\documentclass[a4paper]{spie}  

\usepackage{amsmath,amsfonts,amssymb}
\usepackage[pdftex]{graphicx}
\usepackage[colorlinks=true, allcolors=blue]{hyperref}
\usepackage{url}

\title{Proxy System with JPEG Bitstream-Based File-Size Preserving Encryption for Cloud Photo Streams}

\author[a]{Hiroyuki KOBAYASHI}
\author[b]{Hitoshi KIYA}
\affil[a]{Tokyo Metropolitan College of Industrial Technology, Shinagawa-ku, Tokyo, Japan}
\affil[b]{Tokyo Metropolitan University, Hino-shi, Tokyo, Japan}

\authorinfo{Further author information: \\Hiroyuki KOBAYASHI: E-mail: hkob@metro-cit.ac.jp\\  Hitoshi KIYA: E-mail: kiya@tmu.ac.jp}

\newcommand{\wfig}[1]{Fig. \ref{fig:#1}}
\newcommand{\wfigure}[1]{Figure \ref{fig:#1}}

\newcommand{\wtable}[1]{Table \ref{tab:#1}}

\begin{document}
\maketitle

\begin{abstract}
  In this paper, we propose a proxy system with JPEG bitstream-based file-size preserving encryption to securely store compressed images in cloud environments.
  The proposed system, which is settled between client's device and the Internet, allows us not only to have exact the same file size as that of original JPEG streams but also to maintain a predetermined image format.
  In an experiment, the proposed system is verified to be effective in two cloud photo steams: Google Photo and iCloud Photo.
\end{abstract}


\section{Introduction}

The rapid spread of cloud environments has made using cloud storages and photo streams easy.
However, cloud providers are known to be untrusted, so uploaded images are required to be stored securely in cloud environments\cite{chuman2018security,Ito2021}.
Full encryption with provable security (like RSA and AES) cannot be applied to this issue, because uploaded images have to meet a predetermined image format such as the JPEG standard, which cloud services require in general.
Several bitstream-based encryption methods have been proposed to solve this issue\cite{Ikeda2011,Kiyab1247217,Unterweger2012,Niu2008,Cheng2016}.

For JPEG 2000 images, bitstream-based encryption methods can maintain the same file size before and after encryption by considering the generation of special marker codes\cite{Ikeda2011,Kiyab1247217}.
In contrast, for JPEG images, although bitstream-based encryption methods have also been proposed, but the convention encryption methods cannot maintain the same file size as that of original bitstreams due to the occurrence or disappearance of JPEG marker codes\cite{Unterweger2012,Niu2008,Cheng2016}, except the method that we proposed\cite{Kobayashi2018,2018JDT0004}.
In those methods, in order to hold the JPEG format, encryption is carried out on the basis of byte stuffing operations, which accidentally generate markers.
As a result, the same file size cannot be maintained before and after encryption.
The change of file sizes caused by the encryption methods makes the implementation of proxy systems difficult, even when the JPEG format is maintained.

Accordingly, we propose an encryption/description proxy system for cloud photo streams, in which an JPEG file-size preserving encryption method\cite{Kobayashi2018,2018JDT0004} is used.
The proposed system, which is settled between the client's device and the Internet, allows us not only to have exact the same file size as that of original JPEG streams but also to maintain a predetermined image format. In an experiment, the proposed system is applied to two cloud photo steams: Google Photo and iCloud Photo, and it is verified to be effective in such cloud environments.
\section{Proposed proxy system}

\subsection{Overview}

\begin{figure}[tb]
    \centering
    \includegraphics[width=0.4\textwidth]{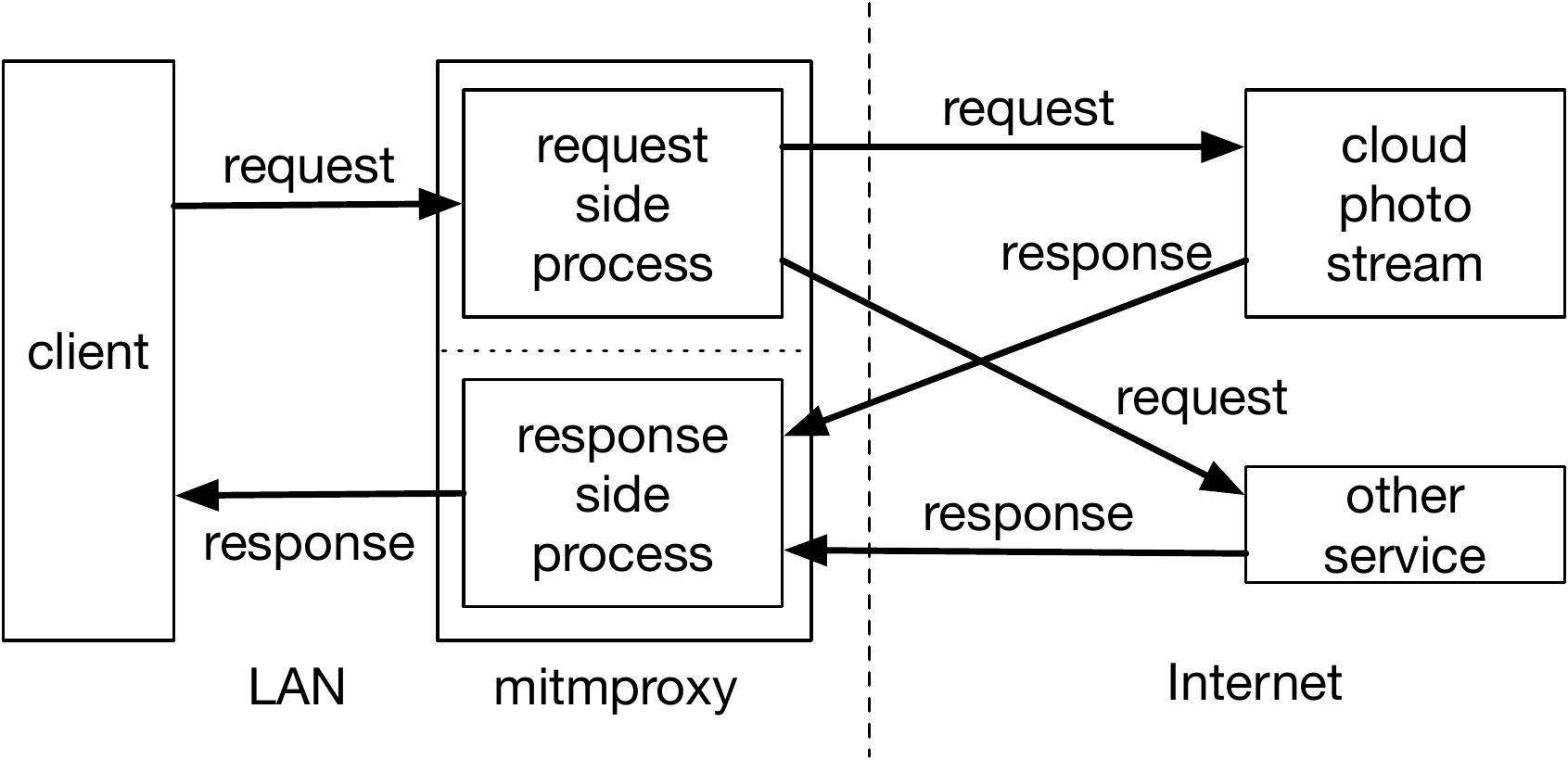} \\[-1mm]
    \caption{Data flow of privacy preserving photo stream}
    \label{fig:proxySystem}
\end{figure}

\begin{figure}[tb]
    \centering
    \begin{tabular}{cc}
        \includegraphics[width=0.5\textwidth]{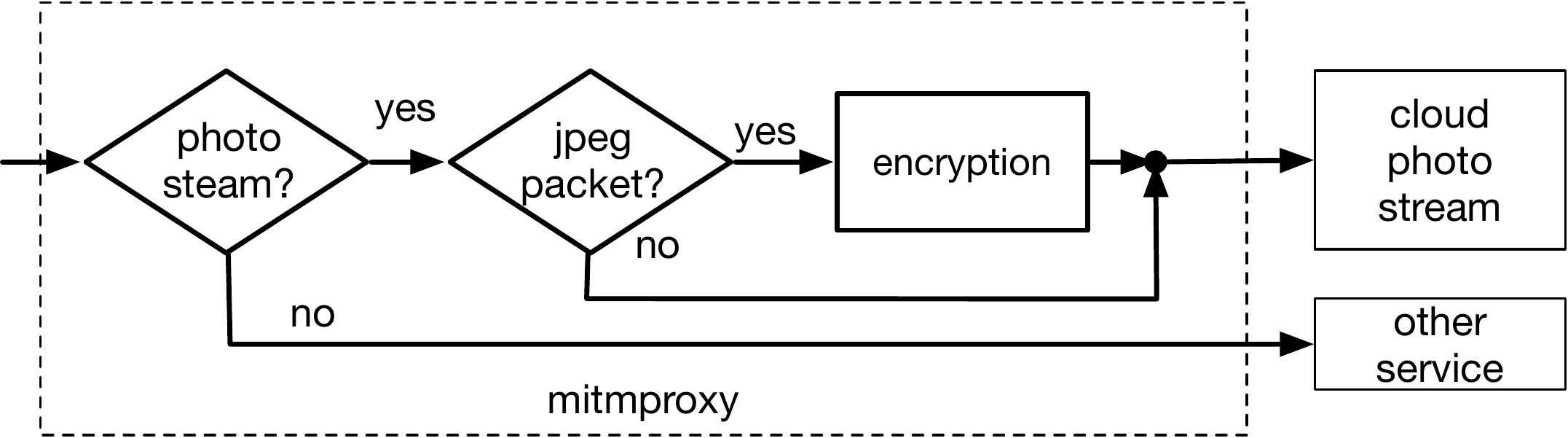} &
        \includegraphics[width=0.45\textwidth]{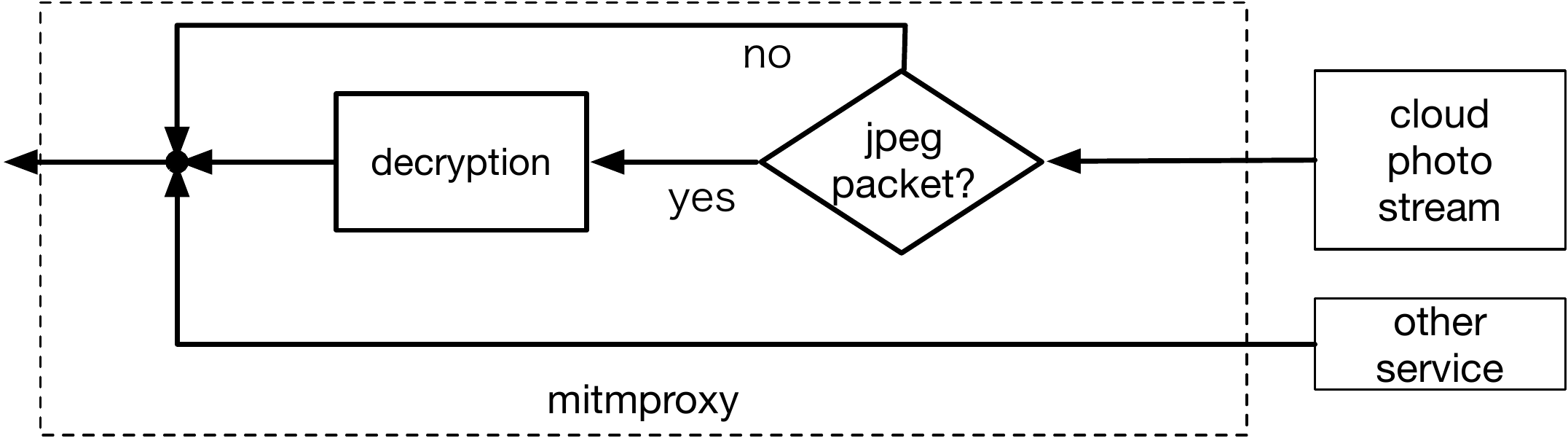}\\
        (a) request side process &
        (b) response side process
    \end{tabular}
    \caption{Proposed proxy system}
    \label{fig:mitmproxy}
\end{figure}

In this paper, an encryption/description proxy system with a JPEG bitstream-based encryption method is proposed for privacy-preserving cloud photo streams.
\wfigure{proxySystem} shows the dataflow of the proxy system implemented with `mitmproxy'\cite{mitmproxy}, where `mitmproxy' is a SSL/TLS-capable intercepting proxy.
The proxy system is designed to confirm whether the destination of each packet is a target cloud photo stream such as Google Photo, and whether each packet includes a JPEG bitstream.
\wfigure{mitmproxy}(a) indicates the procedure of the request side in the proxy system.
JPEG images included in the selected packets are encrypted by using the JPEG bitstream-based encryption\cite{Kobayashi2018,2018JDT0004}.
\wfigure{mitmproxy}(b) shows the procedure of the response side in the proxy system.
A procedure similar to the request side is prepared for the response side.

\subsection{Requirement}
The proposed system satisfies two requirements under the use of encryption: maintaining both the JPEG format and exact the same file size.
If the file size of the JPEG images is changed by encryption, header information on the packet such as `content-length' must also be replaced.
Furthermore, some client applications or photo streams require the same file size.
To apply the proxy system to all environments, it has to meet the two requirements.

\subsection{Encryption with file-size preserving}

\begin{figure}[tbp]
  \centering
  \includegraphics[width=3.3in]{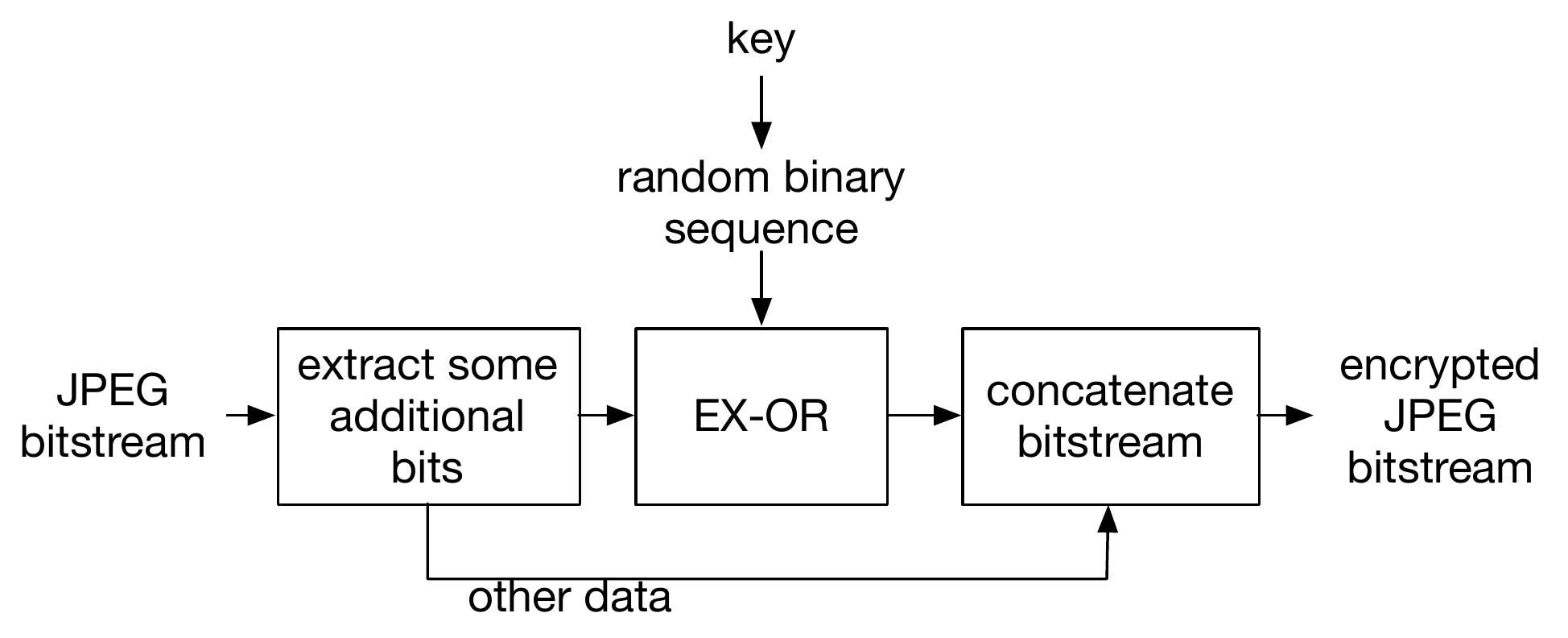}
  \caption{Outline of encryption}
  \label{fig:proposed}
\end{figure}

Encrypted bitstreams have not only the same file sizes, but also the compatibility with JPEG decoders. In order to preserve the file sizes of original bitstreams, the encryption should not change the number of byte stuffing to avoid the generation of markers.
In the encryption method in \cite{Kobayashi2018,2018JDT0004}, a part of additional bits that satisfy some conditions are encrypted to keep the compatibility with JPEG decoders.
\wfigure{proposed} illustrates the outline of the encryption.
The procedure of encryption is summarized as follows.
\begin{enumerate}
  \item Analysis, byte-by-byte, the entropy-coded data segment and extract additional bits from a byte that satisfies two conditions: the byte includes both Huffman code and additional bits, and the Huffman code includes at least one ``0'' bit.
  \item Generate a random binary sequence with a secret key.
  \item Carry out exclusive-or operation between only extracted additional bits and the random sequence generated in 2, and replace the additional bits with the result.
  \item Produce an encrypted bitstream by combining the encrypted additional bits with other data without any encryption.
\end{enumerate}
By analyzing the Huffman code in the byte as described above, it is possible to determine whether encryption is possible or not. The following bytes are not encrypted.
\begin{enumerate}
  \item The whole 8 bits are Huffman codes.
  \item The whole 8 bits are additional bits.
  \item ``00'' byte immediately after ``FF''
  \item Huffman code and additional bits are included and the all bits of Huffman code are `1'.
\end{enumerate}

As a result, the encryption guarantees a constant file size by providing a mechanism to avoid occurrences / disappearances of the byte stuffing.
The encrypted packet has exactly the same size as that of the original packet, so the proxy system is available for any application without any modification of the header information of the packet.


\section{Experiment}
We evaluated the proposed proxy system under two cloud photo streams; Google Photo and iCloud Photo.
The proposed system was settled on LAN and the IP address of the system was set in user's client device.
This proxy system encrypts and replaces only the JPEG images that are transmitted to Google Photo and iCloud Photo.
In this system, we used a JPEG bitstream-based image encryption where the encryption can be carried out in the bitstream domain\cite{Kobayashi2018}, and the additional bits information in both DC and AC components were encrypted.
Since the encrypted image files have exactly the same size as that of original JPEG bitstreams, the proxy system does not need other information such as the content size of the transmit packets.

\subsection{Image Quality Evaluation}

\begin{figure}[htbp]
    \centering
    \begin{tabular}{ccc}
    \includegraphics[width=0.2\textwidth]{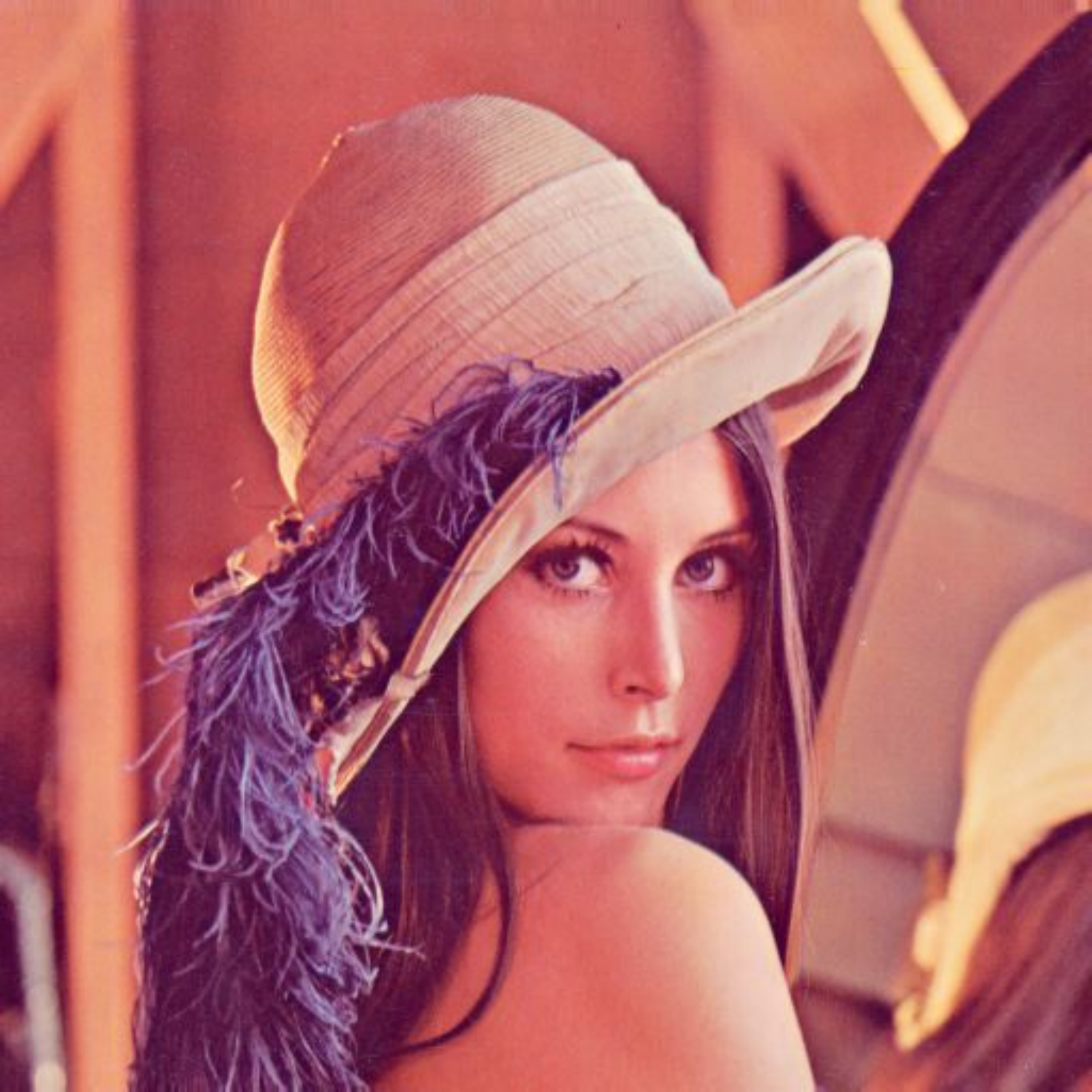}&
    \includegraphics[width=0.2\textwidth]{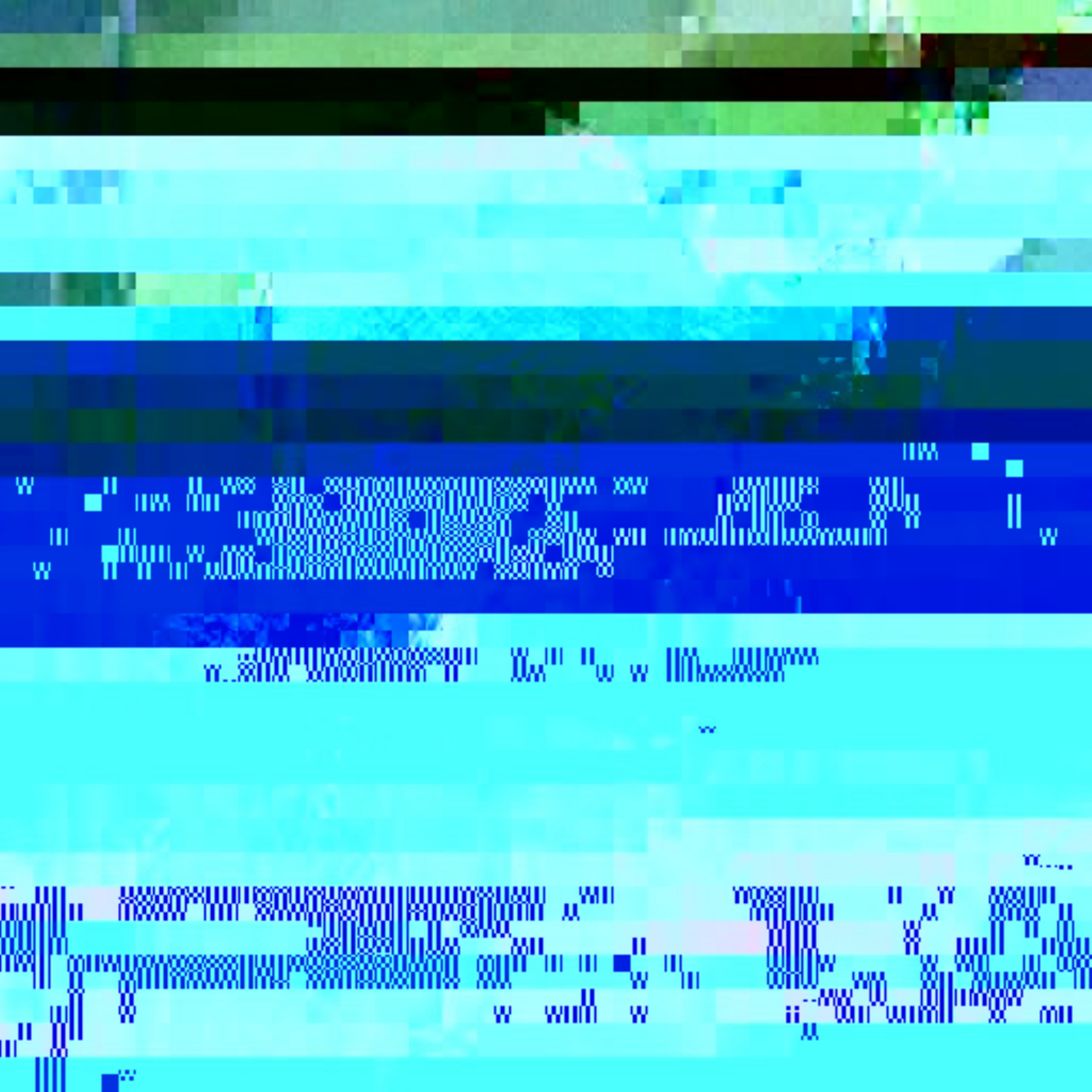} &
    \includegraphics[width=0.05\textwidth]{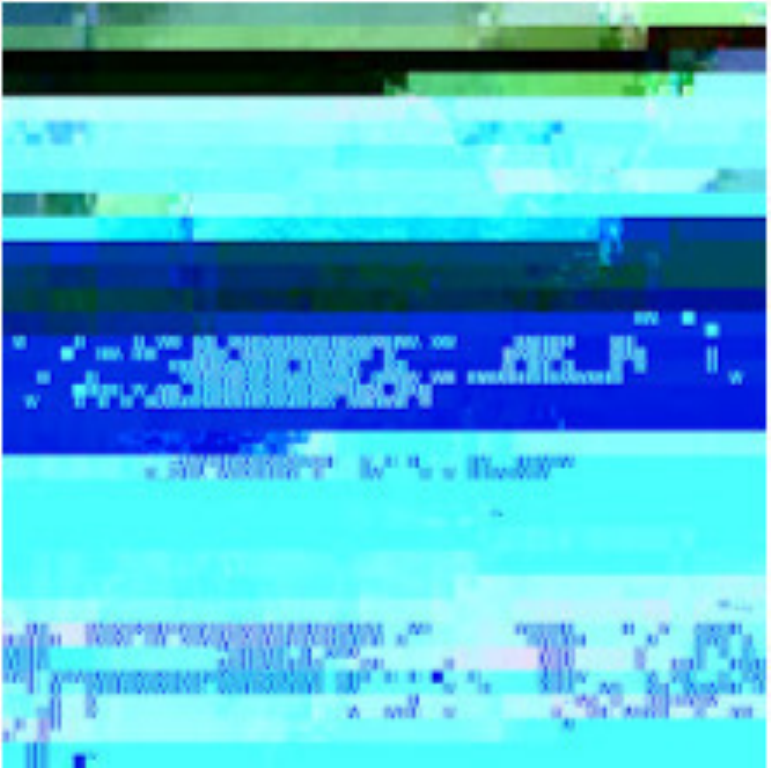} \\
    (a) plain (lena) &
    (b) encrypted image (lena) &
    (c) thumbnail (lena, server) \\
    \includegraphics[width=0.2\textwidth]{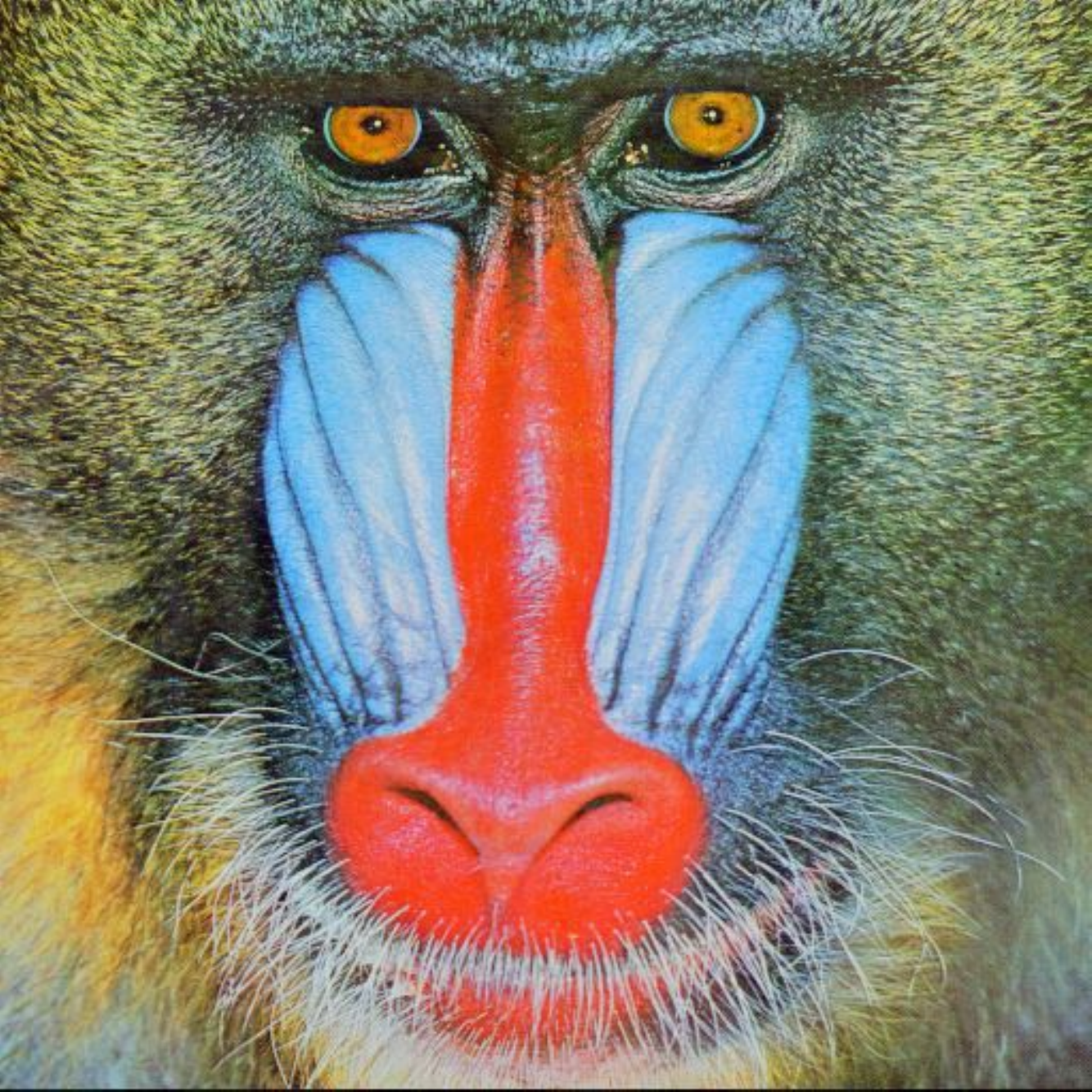}&
    \includegraphics[width=0.2\textwidth]{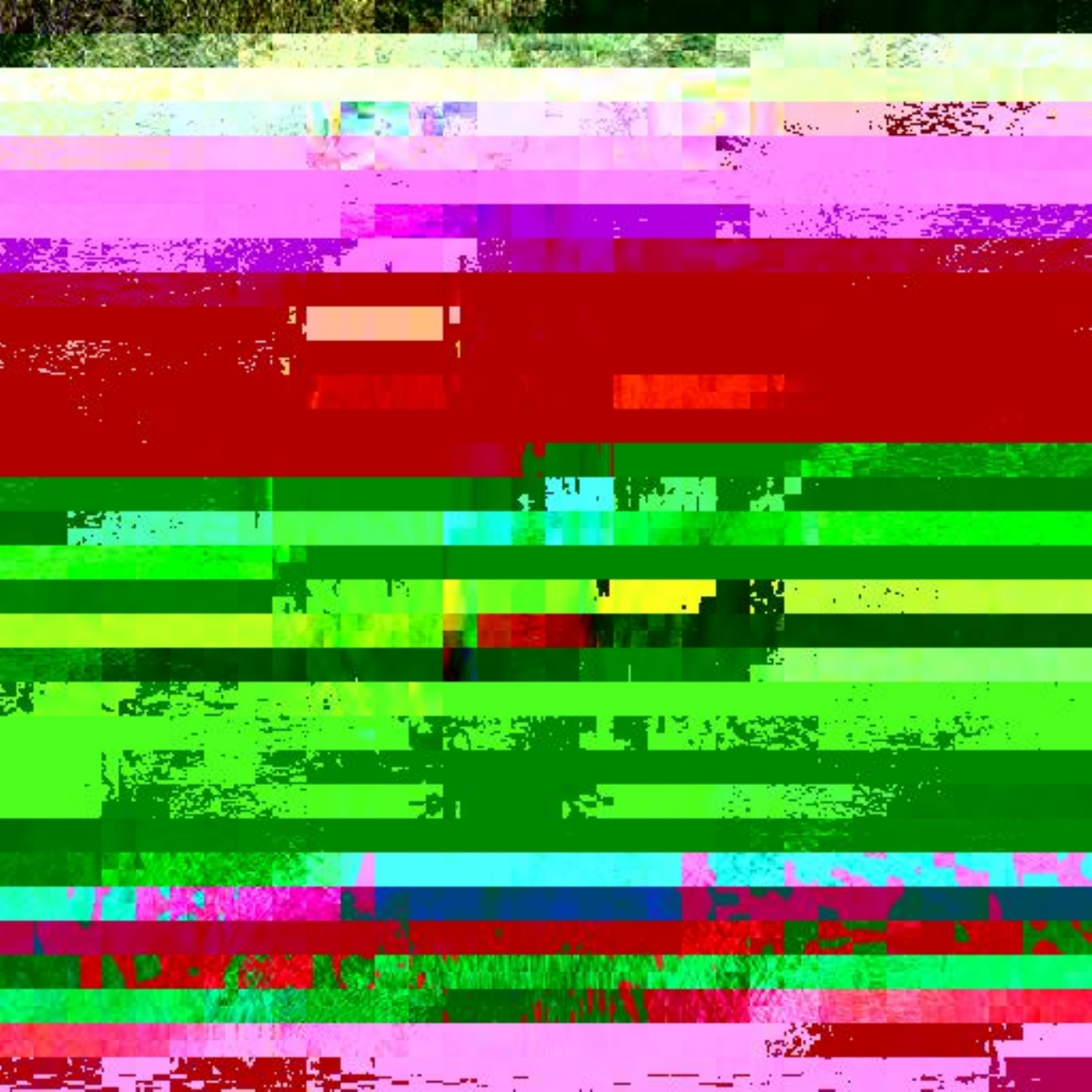} &
    \includegraphics[width=0.05\textwidth]{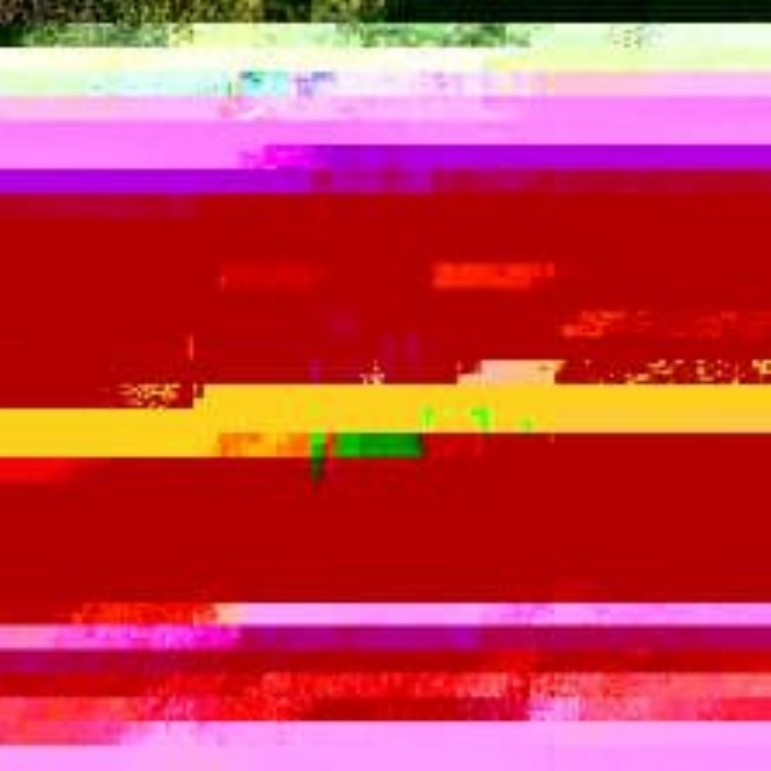} \\
    (d) plain (mandrill) &
    (e) encrypted image (mandrill) &
    (f) thumbnail (mandrill, server)
    \end{tabular}
    \caption{Example of encrypted and thumbnail images}
    \label{fig:encrypedImage}
\end{figure}

Figure \ref{fig:encrypedImage} shows an example of encrypted and thumbnail images.
The plain image(\wfig{encrypedImage}(a), \ref{fig:encrypedImage}(d)) was encrypted by the proxy system with key K, and the encrypted image(\wfig{encrypedImage}(b), \ref{fig:encrypedImage}(e)) was transmitted and registered on the server.
Authorized users with key K downloaded the decrypted files by key K via the proxy.
Figure \ref{fig:encrypedImage}(c), \ref{fig:encrypedImage}(f) shows a thumbnail image registered on the server.
Although thumbnail images were created from the encrypted images by the cloud provider, they was confirmed to still have no visual information on plain images.


\subsection{File Size Preserving}

Next, we compare our encryption method with the previous works\cite{Cheng2016,KURIHARA2015}, in terms of the file sizes.
\wtable{changeSize} shows the file sizes of encrypted JPEG images under various conditions.
From this table, JPEG images encrypted by the proposed method had exactly the same file sizes as those of the original ones.
However, other encryption methods could not preserve the same file sizes, because they do not consider the effect of byte stuffing.


\begin{table}[tb]
  \centering
  \caption{Length of original and encrypted images (difference)[byte], for lena image}
  \label{tab:changeSize}
  \begin{tabular}{c||c|c}
  Q-factor & 80 & 95 \\ \hline\hline
  Original & 43,879 & 106,548 \\ \hline
  \textbf{Proposed} & \textbf{43,879(0)} & \textbf{106,548(0)} \\ \hline
  Cheng\cite{Cheng2016} & 43,865(-14) & 106,553(+5) \\ \hline
  EtC\cite{KURIHARA2015} & 44,487(+608) & 108,262(+1,714) \\
  \end{tabular}
  \end{table}


\section{Conclusion}
We proposed an encryption/description proxy system for cloud photo streams.
The clients could automatically store encrypted images to the cloud photo streams simply by setting the proxy server.
The proposed system was confirmed to work well for Google Photo and iCloud.

\bibliography{library}
\bibliographystyle{spiebib} 

\end{document}